\documentclass{raa_twocolumn}            

\usepackage{graphicx,amsmath,amssymb}             

\begin{document}

   \title{On the nature of long period radio pulsar GPM J1839$-$10: death line and pulse width
}

   \volnopage{Vol.0 (20xx) No.0, 000--000}      
   \setcounter{page}{1}          

   \author{H. Tong}

   \institute{School of Physics and Materials Science, Guangzhou University, Guangzhou 510006, China;
   {\it tonghao@gzhu.edu.cn}\\
   }

   \date{Received~~2009 month day; accepted~~2009~~month day}

\abstract{Recently another long period radio pulsar GPM J1839$-$10 is reported, similar to GLEAM-X J162759.5$-$523504.3. Previously, the energy budget and rotational evolution of long period radio pulsars had been considered.  This time, the death line and pulse width for neutron star and white dwarf pulsars are investigated. The pulse width is included as the second criterion for neutron star and white dwarfs pulsars. It is found that: (1) PSR J0250+5854 and PSR J0901$-$4046 etc should be normal radio pulsars. They have narrow pulse width and they lie near the radio emission death line.  (2) The two  long period radio pulsars GLEAM-X J162759.5$-$523504.3 and GPM J1839$-$10 is unlikely to be normal radio pulsars. Their possible pulse width is relatively large. And they lie far below the fiducial death line on the $P-\dot{P}$ diagram. (3)  GLEAM-X J162759.5$-$523504.3 and GPM J1839$-$10 may be magnetars or white dwarf radio pulsars. At present, there are many parameters and   uncertainties in both of these two possibilities.
}

   \authorrunning{H. Tong}            
   \titlerunning{One the nature of GPM J1839$-$10}  

   \maketitle

%
%
\section{Introduction}
Pulsars are rotating magnetized neutron stars (Lyne \& Graham-Smith 2012). Their rotation period can range from milliseconds to that of several seconds (Young et al. 1999). The period of magnetars can be as large as 12 seconds (Olausen \& Kaspi 2014). Recently, there are new discoveries toward the long period end of pulsars. PSR J0250+5854 is the slowest radio pulsar at that time, with a period of $23.5 \ \rm s$ (Tan et al. 2018). Later more radio pulsars with periods larger than 10 seconds are discovered (Han et al. 2021). The rotational evolution of these long period radio pulsars may evolve physics in the magnetosphere or magnetic field decay (Kou et al. 2019). The central compact object inside supernova remnant RCW 103 showed magentar activities and was confirmed to be a magnetar with a period about 6.6 hours (D'Ai et al. 2016; Rea et al. 2016). Its rotational evolution may require a magnetar spin-down by a fallback disk (Tong et al. 2016). However, the magnetar inside RCW 103 is radio quiet. It is not sure why it is radio quiet, e.g. beaming or no radio emission at all.

The long period radio pulsar (dubbed as LPRP) is push forward by the discovery of GLEAM-X J162759.5$-$523504.3 (GLEAM-X J1627 for short, Hurley-Walker et al. 2022).  GLEAM-X J1627 has a period of $1091 \ \rm s$ (about 18 minutes) and it is a transient in radio. The existence of LPRPs  is further strengthened by PSR J0901$-$4046 with a period of  $76 \ \rm s$ (Caleb et al. 2022). However, it is not certain whether these LPRPs  are neutron stars or white dwarfs. For such slow rotators, it is more natural that they are white dwarfs (Katz 2022; Loeb \& Maoz 2022). At the same time, similar to the case of RCW 103 magnetar, these LPRPs  may also be magnetar+fallback disk systems (Ronchi et al. 2022; Tong 2023). At present, their nature remains unclear (Rea et al. 2022).

A source similar to GLEAM-X J1627 is recently reported, GPM J1839$-$10 (Hurley-Walker et al. 2023). GPM J1839$-$10 has a pulsation period of $1318 \ \rm s$ (about 21 minutes). Compared with GLEAM-X J1627, GPM J1839$-$10 has (1) a more stringent period derivative upper limit $\dot{P} < 3.6\times 10^{-13}\ \rm s \ s^{-1}$ (Hurley-Walker et al. 2023), which can place more constraint on the pulsar death line. (2) its radio emission has lasted for more than 30 years. This can enable later more monitoring.

For the energy budge and rotational evolution of GPM J1839$-$10, similar calculations can be applied to it as that for GLEAM-X J1627 (Katz 2022; Loeb \& Maoz 2022; Ronchi et al. 2022; Tong 2023). Since their periods are similar (21 minutes verse 18 minutes), the same conclusion may also be applied to GPM J1839$-$10, i.e., magnetar+fallback disks or white dwarf pulsars. We will not repeat the calculations here. This time, we focus on (1) the death line for neutron  star and white dwarf pulsars, (2) pulse width of LPRPs. From these two aspects, we want to discuss the nature of GPM J1839$-$10, which is the most recent example of LPRPs.

\section{Model calculations}

\subsection{Death line for neutron star and white dwarf pulsars}

The death line for radio emission of normal pulsars and magnetars has been discussed in Section 2.5 in Tong (2023). The potential drop at a specific angle across the polar cap is just a classical electrodynamics exercise. For normal radio pulsars with a dipole magnetic field, the maximum acceleration potential across the polar cap is (Ruderman \& Sutherland 1975; Zhou et al. 2017):
\begin{equation}\label{eqn_Phimax}
\Phi_{\rm max} = \frac{B_p R^3 \Omega^2}{2c^2} \equiv 10^{12} \ \rm V,
\end{equation}
where $B_p$ is the surface magnetic field at the pole region (which is two times the equatorial value, Lyne \& Graham-Smith 2012), $R$ is the stellar radius, $\Omega$ is the star's angular velocity. When the maximum acceleration potential equals $10^{12} \ \rm V$, it  is defined as the radio emission death line. Below the death line, the star is not expected to have radio emission. The value of $10^{12} \ \rm V$ is just a fiducial value. A rough estimation of the physics evolved is that (Ruderman \& Sutherland 1975): an electron accelerated in such a potential attains a Lorentz factor about $\gamma \sim 10^6$. This electron may emit curvature photons of energy
\begin{equation}\label{eqn_curvature_photon_energy}
h \nu = h \frac{3\gamma^3 c}{4\pi \rho} \ge 1 \ \rm MeV.
\end{equation}
These curvature photons may be converted to electron-position pairs in strong magnetic fields. If the electron Lorentz factor (i.e., acceleration potential) is lower, the curvature photons may have energy less than $1 \ \rm MeV$. The subsequent pair production process can not continue. And this may result in cease of the radio emission (i.e., radio emission death line).

For  magnetar+fallback disk system, the magnetosphere may be modified by (1) the fallback disk if the disk is still active, (2) the magnetar's twisted magnetic fields. If the death line is modified by the fallback disk, the corotation radius will replace the light cylinder radius as the maximum radial extent of the field lines. The corresponding maximum acceleration potential and death line is presented in Eq. (10) in Tong (2023):
\begin{equation}
\Phi_{\rm max,disk} = \frac{B_p R^3 \Omega^2}{2c^2} \frac{R_{lc}}{R_{co}} \equiv 10^{12} \ \rm V,
\end{equation}
where $R_{lc} = Pc/(2\pi)$ is  the neutron star's light cylinder radius, $R_{co} = (G M/4\pi^2)^{1/3} P^{2/3}$ is the corotation radius.
If the death line is modified by the twist of the field lines, a twisted  field line will result in a larger polar cap and a larger potential drop. The maximum acceleration potential and death line for a twisted magnetic field is presented in Eq. (12) in Tong (2023):
\begin{equation}
\Phi_{\rm max,twist} = \frac{B_p R^3 \Omega^2}{2c^2} \left( \frac{R_{lc}}{R} \right)^{1-n} \equiv 10^{12} \ \rm V,
\end{equation}
where $n$ is a parameter characterizing the twist of field lines. $n=1$ corresponds to the dipole case (for $n=1$, the equation returns to the dipole case Eq.(\ref{eqn_Phimax})). $0<n<1$ corresponds to the twisted dipole case.

For white dwarf pulsars, it is possible that they have similar magnetospheric precess to that of neutron star pulsars (Zhang \& Gil 2005; Katz et al. 2022). However, there are several changes in the definition of death line (Eq.\ref{eqn_Phimax}) for white dwarf pulsars.
\begin{enumerate}
\item For a typical neutron star, the radius is usually set to be $10 \ \rm km$. For a typical white dwarf (figure 5.17 in Camenzind 2007), it has a  radius of one precent the solar radius $0.01\ R_{\odot}$ (the corresponding white dwarf mass is about $0.8 \ M_{\odot}$).

\item The torque of a rotating magnetized object can be approximated as magnetic dipole braking (Xu  \& Qiao 2001). The star's magnetic field can be obtained from the period and period-derivative measurement, which is a crude estimate of the star's true magnetic field. Assuming a perpendicular rotator, the magnetic field is
\begin{equation}
B = \sqrt{\frac{3I c^3}{8\pi^2 R^6} P \dot{P}},
\end{equation}
where $B$ the equatorial magnetic field at the surface (it is two times smaller than the polar magnetic field), $I$ is the star's moment of inertia. For typical neutron stars, with $I \approx 10^{45} \ \rm g \ cm^2$, $R\approx 10^6 \ \rm cm$, it is the commonly cited formula for radio pulsars:
\begin{equation}
B = 3.2 \times 10^{19} \sqrt{P \dot{P}} \ \rm G.
\end{equation}
For white dwarfs, assuming a typical mass of $0.8 \  M_{\odot}$ and radius of $0.01 \ R_{\odot}$, the white dwarf's moment of inertia is about $3 \times 10^{50} \ \rm g \ cm^2$. Therefore, the characteristic magnetic field for a white dwarf pulsar is:
\begin{equation}
B = 5.2 \times 10^{13} \sqrt{P \dot{P}} \ \rm G.
\end{equation}
This formula will be employed when drawing the death line on the $P-{\dot P}$ diagram.

\item From Eq. (\ref{eqn_curvature_photon_energy}), a minimum Lorentz factor (i.e. acceleration potential) is require to generate curvature photons with energy higher than $1 \ \rm MeV$. In the case of white dwarfs, the stellar radius is larger. The curvature radius of the magnetic field line is also larger, which is order of $\sqrt{r R_{lc}}$ ($r$ is the emission height, Xu \& Qiao 2001). Therefore, a higher acceleration potential may be required, e.g. as high as $10^{13} \ \rm V$ in Eq. (\ref{eqn_Phimax}). The exact value depends on the detailed modeling  of the white dwarf's magnetosphere (as in the case of neutron star pulsars).

\end{enumerate}

The death line for normal radio pulsars, magnetars and magnetar+fallback disk systems is shown in figure 1, along with GPM J1839$-$10 and other LPRPs. Figure 1 is updated from figure 2 in Tong (2023). The death line for white dwarf pulsars is  so different from that of neutron star case that it is shown separately in figure 2. $\Phi_{\rm max} =  10^{12} \ \rm V$ and $\Phi_{\rm max} = 10^{13} \ \rm V$ are shown respectively. The characteristic magnetic field for white  dwarf pulsars is also shown, for $B=10^8 \ \rm G$  and $B=10^9 \ \rm G$. Most of the presently observed pulsating white dwarfs have magnetic field smaller than $10^9 \ \rm G$ (Zhang \& Gil 2005; Katz 2022; Marsh  et al. 2016; Pelisoli et al. 2023).

\begin{figure}[htbp!]
  \centering
  \includegraphics[width=0.475\textwidth]{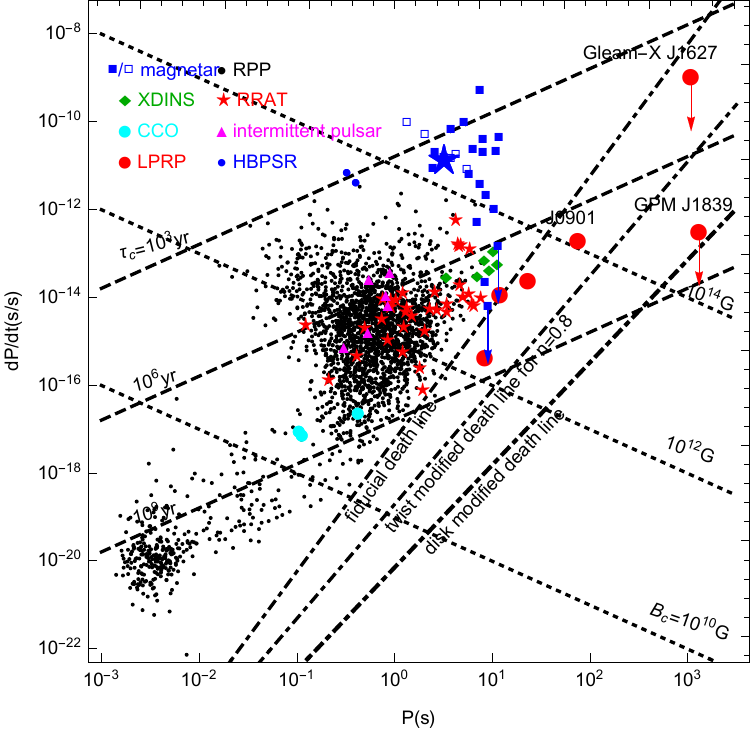}\\
  \caption{Definition of death line and distribution of long period radio pulsars (red circles) on the $P$-$\dot{P}$ diagram. The fiducial pulsar death line, the death line for a twisted magnetic field ($n=0.8$), and the death line modified by the fallback disk are also shown. Updated from figure 2 in Tong (2023). }
  \label{fig_LPRP}
\end{figure}

\begin{figure}[htbp!]
  \centering
  \includegraphics[width=0.475\textwidth]{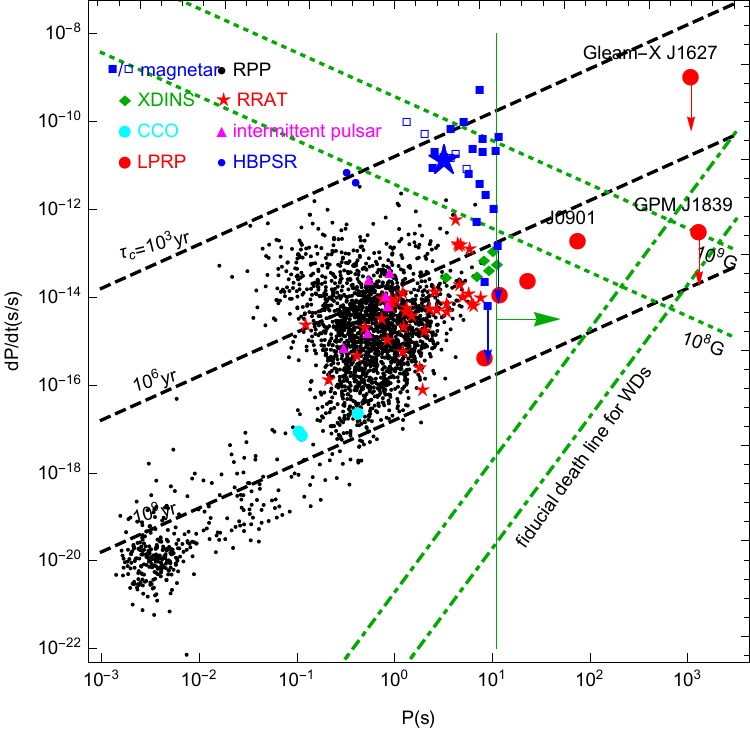}\\
  \caption{Definition of death line and contour of constant magnetic field for white dwarf pulsars. Two death lines are shown, with $\Phi_{\rm max} = 10^{13} \ \rm V$ (upper one) and $\Phi_{\rm max} = 10^{12} \ \rm V$ (lower one) respectively. Two contours of constant magnetic field are shown, with $B=10^9 \ \rm G$ and $B=10^{8} \ \rm G$ respectively. The limiting Keplerian period for a typical white dwarf pulsar is also shown. }
  \label{fig_WD_pulsar}
\end{figure}

\subsection{Pulse width}

GPM J1839$-$10 has a period of $1318 \ \rm s$. The single pulse vary in a pulse window $400 \ \rm s$ (Hurley-Walker et al. 2023). Similar things also happen in normal radio pulsars and radio emitting magnetars (Levin et al. 2012; Yan et al. 2015; Huang et al. 2021). At present, the integrated pulse profile of GMP J1839$-$10 is not available. From previous experiences in pulsars and magnetars, the pulse window may be an estimate of the integrated pulse width. Then the pulse width of GPM J1839$-$10 is $PW \approx 400/1318 = 30\%$ of the pulse phase (in this case, the pulse width may also be called the duty cycle, Tan et al. 2018). If this is the pulse width of GPM J1839$-$10, it can also constrain the nature of the source.

For normal radio pulsars, the colatitude of the last open field line at emission height $r$ is:
\begin{equation}
\theta_{\rm open} = \sin^{-1} \left(  \frac{r}{R_{lc}} \right)^{1/2}.
\end{equation}
The emission beam radius is $3/2$ times the angle $\theta_{\rm open}$ (Eq.(17.9) in Lyne \& Graham-Smith 2012):
\begin{equation}\label{eqn_rho_beam}
\rho_{\rm beam} = \frac32  \sin^{-1} \left(  \frac{r}{R_{lc}} \right)^{1/2}.
\end{equation}
The emission height of normal radio pulsars is always less than $100$ times  the neutron star radius (Johnston et al. 2023). Therefore, the emission height in Eq.(\ref{eqn_rho_beam}) can be set as $100 R$, where $R$ is the stellar radius.
The observed pulse width depends also on the inclination angle $\alpha$ (angle between rotation axis and line of sight), and impact angle $\beta$ (closest approach between the magnetic axis and the line of sight). The impact angle may always be a small quantity, in this case the observed pulse width is related to the emission beam radius as (Eq.(15.2) in Lyne \& Graham-Smith 2012):
\begin{equation}
W= 2\rho_{\rm beam} \frac{1}{\sin\alpha}.
\end{equation}
In terms of pulse phase, the observed pulse width is:
\begin{equation}\label{eqn_PW_dipole}
PW = \frac{W}{2\pi} = \frac{3}{2\pi} \sin^{-1} \left(  \frac{r}{R_{lc}} \right)^{1/2} \frac{1}{\sin\alpha}.
\end{equation}
A small  inclination angle $\alpha$ can result in a large pulse width.

For a magnetar+fallback disk system, the calculation is similar to the calculations for death line. If the magnetosphere is regulated by the fallback  disk, the corotation radius replaces the role of light cylinder radius in Eq. (\ref{eqn_PW_dipole}). For a twisted magnetic field, the  colatitude of the last open field line will be larger (Eq. 11 in Tong 2023). The pulse width in units of pulse phase is:
\begin{equation}
PW = \frac{3}{2\pi} \sin^{-1} \left(  \frac{r}{R_{lc}} \right)^{n/2} \frac{1}{\sin\alpha}.
\end{equation}
For $n=1$, the above equation returns to the dipole case (Eq.(\ref{eqn_PW_dipole})).
For a white dwarf pulsar, the expression for the pulse width is the same as Eq. (\ref{eqn_PW_dipole}), except that the stellar radius should be the white dwarf radius.

The theoretical pulse width for normal pulsars, magnetar+fallback disk systems, and white dwarf pulsars is shown in figure 3. Magnetars and white dwarfs pulsars have wider pulse width compared with that of normal radio pulsars. This is especially true for LPRPs.

\begin{figure}[htbp!]
  \centering
  \includegraphics[width=0.475\textwidth]{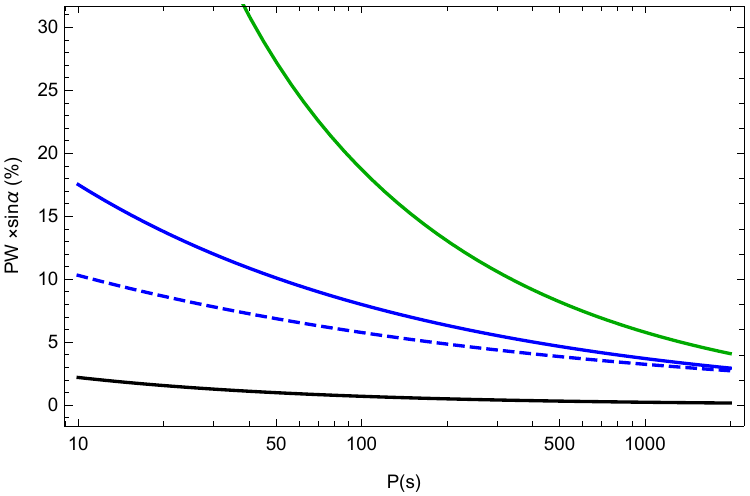}\\
  \caption{Theoretical pulse width as a function of period. The pulse width is in units of pulse phase. From bottom to top are: normal radio pulsars (black), magnetars with twisted magnetic field (dashed blue, for $n=0.5$), magnetar+fallback disk systems (solid blue), and white dwarf pulsars (green). Since the inclination angle is unknown, the plotted pulse width is actually $PW \times \sin\alpha$. For a small $\alpha$, the actual pulse width can be larger.}
  \label{fig_PW}
\end{figure}

\subsection{On the nature of GPM J1839$-$10}

From the death line on the $P-\dot{P}$ diagram, the LPRP GPM J1839$-$10 can not be normal radio pulsars. A twisted magnetic field or the presence of a fallback disk can help to lower the position of the death line on the $P-\dot{P}$ diagram. The quantitative result depends on the parameters involved, e.g. the twist parameter $n$. Figure 1 shows the death line for a twisted magnetic field with $n=0.8$. For a more twisted magnetic field (i.e. $n=0.5$), the position of the death line is lower on the $P-\dot{P}$ diagram. However, the typical untwisting timescale may be smaller, which may have difficulties in explaining why GPM J1839$-$10 can have radio emission lasting more than 30 years. A magnetar+fallback disk system may be consistent with the position of GPM J1839$-$10 on the $P-\dot{P}$ diagram. However, the generation of radio emission in the presence of an active fallback disk is uncertain, although there are such possibilities (see Section 2.5 in Tong 2023  for discussions).

The death line for white dwarf pulsars is also consistent with the position  of GPM J1839$-$10 on the $P-\dot{P}$ diagram. However, there may be three constrains for white dwarf pulsars: (1) The maximum acceleration potential for white dwarf pulsars may be higher, e.g., as high as $10^{13} \ \rm V$. (2) White dwarf pulsars generally have magnetic field lower than $10^9 \ \rm G$ (Zhang \& Gil 2005; Katz 2022; Marsh  et al. 2016; Pelisoli et al. 2023). (3) For a white dwarf with mass $0.8 M_\odot$ and radius $0.01 R_\odot$, the limiting Keplerian period is:
\begin{equation}
P_K = 2\pi \sqrt{\frac{R^3}{G M}} \approx 11 \ \rm s.
\end{equation}
These three constrains will limit the existence of white dwarf pulsars to a small triangle on the $P-\dot{P}$ diagram, see Figure 2. Such a triangle of parameter space may explain why there are so few white dwarf radio pulsars (LPRPs are just candidates).

The pulse with of GPM J1839$-$10 is unknown at present. If its pulse window represents its pulse width, then it will have a pulse width of $30\%$. From the theoretical pulse width, normal radio pulsars may have difficulties in explaining the pulse width of GPM J1839$-$10. The solution in the normal radio pulsar case is that they should have an extremely large emission height. For a twisted magnetic field or magnetar+fallback disk system, the pulse width is larger, typically several precent. A small inclination angle or a slightly higher emission height may explain the possible observed pulse width of GPM J1839$-$10. White dwarf pulsars can naturally have larger pulse width.

By combining the death line and pulse width requirement, it is unlikely that GPM J1839$-$10 is a normal radio pulsar. It is possible that GPM J1839$-$10 is a magnetar (including magnetar+fallback disk system) or a white dwarf pulsar. For these two possibilities, there are many parameters at present. This conclusion is consistent with previous calculations for GLEAM-X J1627, based on energy budget and rotational evolution (Katz 2022; Loeb \& Maoz 2022; Ronchi et al. 2022; Tong 2023). Population synthesis of neutron star and white dwarf pulsars also get similar conclusions (Rea et al. 2023).

\section{Discussion}

{\it Comparison with other LPRPs.} All the three LPRPs PSR J2144-3933 (Young et al. 1999), PSR J0250+5854  (Tan et al. 2018) and PSR J0901$-$4046 (Caleb et al. 2022) have a very narrow pulse width, typically less than $1\%$. Assuming a maximum emission height of $100 R$ (Johnston et al. 2023), the theoretical upper limit on pulse width is (Eq.(\ref{eqn_PW_dipole})):
\begin{equation}
PW < 7\% P^{-1/2} \frac{1}{\sin\alpha}.
\end{equation}
Therefore, PSR J0250+5854 (with a period of $23.5 \ \rm s$) and PSR J0901$-$4046 (with a period of $76 \ \rm s$) etc is consistent with a normal radio pulsar origin, from the pulse width point of view. However, both of GLEAM-X J1627 and GPM J1839$-$10 (Hurley-Walker et al. 2022, 2023) showed possible signatures of a large pulse width. Therefore, we propose that the pulse with of LPRPs may be taken as the second criterion in identifying their nature, in addition to their relative position to the death line on the $P-\dot{P}$ diagram. If a future LPRP has narrow pulse width consistent with that of normal radio pulsars, it may be viewed as an extreme radio pulsar. All we have to do is to consider the corresponding magnetospheric physics involved, e.g. the definition of the death line etc. If a LPRP has a large pulse width, then we must consider the possibility of magnetars  or white dwarf pulsars.

{\it Comparison with other radio emitting magnetars.} Radio emitting magnetars generally have a large pulse width compared with normal radio pulsars (Levin et al. 2012; Yan et al. 2015; Huang et al. 2021). Especially, the single pulse width of magnetars are generally very narrow. They may vary randomly in the pulse window, therefore resulting in a wide integrated pulse width (Levin et al. 2012; Yan et al. 2015). We think this may also be the case of GPM J1839$-$10: narrow single pulse and a wide pulse window. From the pulse width point of view, the two LPRPs GLEAM-X J1627 and GPM J1839$-$10 are similar to radio-emitting magnetars.

{\it Maximum period of LPRPs.} From Figure 1 and 2 (or Figure 1 in Rea et al. 2023), the existence of maximum magnetic field in combination with the death line implies there is a maximum period for radio-emitting neutron stars and white dwarfs. For neutron stars, the maximum magnetic field may be about $10^{16} \ \rm G$. The definition of death line is rather uncertain (which will result in a death valley). However, the possible maximum period for radio emission may be around $10^4 \ \rm s$ or so. In this respect, the magnetar inside RCW 103 (with a period of 6.6 hours) is not expected to have radio emissions. For white dwarfs, assuming of a maximum magnetic field of $10^9 \ \rm G$, the expected maximum period may be about several thousand seconds. Future LPRPs with longer periods may help to unveil their nature, i.e., neutron stars or white dwarfs.

{\it Comparison with previous works.} Observationally, GPM J1839$-$10 is similar to that of GLEAM-X J1627 (Hurley-Walker et al. 2023). It confirms the existence of LPRPs as sub-population of radio pulsars, which may deserved future more searchings. Theoretically, the modeling for GPM J1839$-$10 may seem similar to that of GLEAM-X J1627. Compared with previous theoretical works (Ronchi et al. 2022; especially Tong 2023), the present work has two outstandings: (1) The death line is extended to the case of magnetar's twisted magnetic field and magnetar+fallback disk system in previous works. Here, the pulsar death line is extended to the white dwarf pulsar case, considering possible physics involved during the definition of the death line. (2) The pulse width is newly included in this work. We propose that it may be taken as the second criterion concerning the nature of LPRPs, in addition to that of timing parameters (e.g., period, period-derivative, death line etc). By combining the rotational evolution, energy budget, death line and pulse with, something is already clear for LPRPs: (1) For PSR J0901$-$4046 (Caleb et al. 2022) etc with period less than $76\ \rm s$, they can be understood in the normal radio pulsar domain. (2) For the two LPRPs GLEAM-X J1627 and GPM J1839$-$10 (Hurley-Walker et al. 2022, 2023), they are unlikeky to be normal radio pulsars. Whether they are magnetars or white dwarf pulsars remain inconclusive at present.

\section{Conclusion}

By considering the four aspects together (rotational evolution and energy budget in previous works, death line and pulse width in this work), it is found that: (1) PSR J0250+5854 (Tan et al. 2018) and PSR J0901$-$4046 (Caleb et al. 2022) should be normal radio pulsars. They have narrow pulse width and they lie near the radio emission death line. Further investigations of their magnetospheric physics is required. (2) The two  LPRPs GLEAM-X J1627 and GPM J1839$-$10 (Hurley-Walker et al. 2022, 2023) is unlikely to be normal radio pulsars. Their possible pulse width is relatively large. And they lie far below the fiducial death line on the $P-\dot{P}$ diagram. (3)   GLEAM-X J1627 and GPM J1839$-$10 may be (a) magnetars with twisted magnetic field or magnetar+fallack disk systems, or (b) white dwarf radio pulsars. At present, there are many uncertainties in both of these two possibilities. More multiwave observations are required in order to tell whether they are magnetars or white dwarfs.

\section*{acknowledgments}
H. Tong would like to thank Dr. Huang Zhi-Peng and Yan Zhen for discussions on pulse width and emission height.
This work is supported by National SKA Program of China (No. 2020SKA0120300) and NSFC (12133004).




\label{lastpage}

\end{document}